# Approximation of Search Times for On-street Parking Based on Supply and Demand

Nir Fulman, Itzhak Benenson

Geosimulation Lab, Department of Geography and Human Environment,
Porter School of the Environment and Earth Sciences, Tel Aviv University

nirfulma@post.tau.ac.il, bennya@post.tau.ac.il

**Abstract**

We propose a method for approximating the probability $p(\tau, n)$ of searching for on-street parking longer than time $\tau$ from the start of a parking search near a given destination n, based on high-resolution maps of parking demand and supply in a city. We verify the method by comparing its outcomes to the estimates obtained with an agent-based model of on-street parking search. As a practical example, we construct maps of cruising time for the Israeli city of Bat Yam, and demonstrate that despite the low overall demand-to-supply ratio of 0.65, excessive demand in the city center results in parking searches of longer than 10 minutes. We discuss the application of the proposed approach for urban planning.

## 1. Introduction: Long parking search is characteristic of every city center

Long parking search time is a perpetual problem of every big city, and quantitative estimation of cruising time is a long-standing challenge for transportation research. Given only a moment's thought, it is clear that a parking search may take very long time or even fail when the parking occupation rate is very high and when, during the period of the search, the number of arriving vehicles is similar to or higher than the number of vehicles departing their parking spots. This imbalance between the occupation rate and the rates of arrivals and departures is typical for the central office/commercial/residential areas of city centers, and results in the notorious estimate of 30% of urban traffic being caused by parking search [Shoup, 1997].

Let us recall that, in reality, urban parking conditions are highly heterogeneous in space and vary in time. Parking demand is defined by the size and use of buildings and other attractions in the area, while supply is defined by the parking capacity of street links and off-street facilities, as well as by parking regulations and pricing. The high spatio-temporal heterogeneity of urban parking availability inspires drivers to hope in vain that they will find a spot "just around the corner." At the same time, drivers want to minimize their walk time and confine their search to the area nearby their destination. The rate of arrivals to an area is affected by the attractiveness of its destinations, such as workplaces, offices, shops and residences; while the rate of departures by the activities of the drivers who are parked in the area, such as working, visiting destinations or residential parking. The state of the area of a driver's parking search reflects the net outcome of the arrivals and departures to that area and its immediate surroundings.

Parking search habits of visitors and residents are a reflection of a city's parking regulations and pricing. Let us consider, for example, Tel Aviv city center, where the overall on-street demand-to-supply ratio is above 100% throughout the entire day. On-street parking is free for Tel Aviv residents and inexpensive for non-residents, significantly cheaper than parking in one of the city's numerous parking facilities. In the short run, the parking choices of the drivers in the Tel Aviv center are either (1) cruise until a free spot is found; (2) use an off-street parking facility that is expensive and possibly far from the destination; (3) park illegally, accepting the risk of a high fine; (4) leave the area. As the last three options are very unattractive for residents parking overnight, they often cruise for a quarter of an hour or more when returning home from work [Levy et al., 2015] and yet their final parking spot may be very far from their home. In the long run, frequent long cruising may result in a change of transportation habits and decrease car ownership levels [Weinberger et al., 2009], but as far as Tel Aviv is concerned, there is not yet any meaningful decrease in the demand-to-supply ratio, and long cruising for on-street parking remains one of major issues of the Tel Aviv municipal elections.

Quantitative understanding of the parking search that accounts for the inherent heterogeneity of the urban parking space thus remains one of the basic issues of urban transportation planning and management. In this paper, we propose a simple algorithm for estimating cruising time curves – the probability p(τ, n) of searching for parking longer than τ for a given destination n in the city. The algorithm is based on high-resolution maps of urban parking demand and supply that are available for the majority of large Western cities. We validate the algorithm with PARKGRID, an agent-based model of urban parking search. As a practical example we construct cruising time curves for the Israeli city of Bat Yam, with a population of 120,000.

## 2. Literature review

Parking search time is defined by the parking occupation pattern around a driver's destination, which varies over time in respect to the arrivals and departures within the area. In order to estimate cruising time curves, therefore, we need a high-resolution in space and in time dynamic model of on-street parking search, that accurately reproduces the observed parking pattern and its fluctuations.

### 2.1. Modeling urban parking search and dynamics of the urban parking pattern

Cruising for curb parking is studied with stochastic or deterministic analytical and simulation models [Arnott and Inci, 2006; Arnott and Rowse, 1999; 2013; Anderson and de Palma, 2004; Levy et al., 2013, Dowling et al., 2017; Geroliminis, 2015]. Deterministic analytical models are, typically, aggregated and focus on the average search time $\hat{t}$ as a function of the average occupation rate $\hat{q}$, arrival rate $\lambda$, and departure rate $\mu$. Stochastic analytical models follow queue theory and consider street links as queue servers, a street network as a network of servers, and cruising cars as joining and changing queues. If a car is rejected by a queue server, it moves to another queue server with respect to the street network topology [Dowling et al., 2018a, 2018b]. These models are able to incorporate the emerging heterogeneity of the curb parking pattern and generate likely estimates of the occupancy of street links [Dowling et al., 2018a], but not, however, of cruising time.

Simulation modeling of parking makes it possible to account for the details of the parking scenario and, thus, to provide likely estimates of two critical policy parameters – the distribution of drivers' search times and the distance between drivers' parking places and their destinations [Levy et al., 2013; Levy and Benenson, 2015]. Simulation models of cruising for parking may account for the car following effect [Levy and Benenson, 2015; Arnott and Williams, 2017], but we are not aware of analytical models that explicitly account for this phenomenon.

### 2.2. Estimating cruising time in models of parking search

Basically, the higher the occupation rate in the search area, the longer the search time. However, the quantitative estimates of this dependency in the analytical models are different from the estimates obtained in dynamic simulations. According to Levy et al. [2013] average cruising time $\hat{t}$ in a grid city characterized by patterns of homogeneous demand and supply is essentially non-zero starting from $\hat{q} \approx$ 90%. At the same time, in their analytical model $\hat{t}$ remains insignificant up to a very high $\hat{q}$ of ca. 98%. Levy et al. [2013] explain this gap by the emerging clusters of fully occupied parking spots that cannot be captured by an analytical model operating with averages. Arnott and Williams [2017] term this phenomenon *the bunching effect*: Occupied parking spots tend to "bunch up" into spatially contiguous clusters because drivers traverse consecutive spots until they find a vacant one. Due to the clustered pattern of occupied parking spots, the estimates of $\hat{t}$ obtained for the uniform parking pattern are always lower than those for the heterogeneous one, and this discrepancy increases with the increase in $\hat{q}$. To properly estimate cruising time, we must explicitly account for the inherent spatio-temporal variation in the parking pattern that is defined by the heterogeneity of the demand and supply on the one hand and the stochastic nature of the parking process on the other.

### 2.3. Modeling parking search behavior in dynamic simulations

Simulation of parking search demands explicit representation of drivers' cruising behavior: their attitudes to parking regulations, their ability to assess current parking occupation rates within their search area, and their ability to learn from experience. All of the simulation models we are aware of employ heuristic rules of drivers' cruising behavior and decision making when selecting a vacant spot. In the MATSim's implementation, the cruising path is constructed as a random sequence of connected links that start from a driver's destination [Bischoff and Nagel, 2017]. In the PARKAGENT model [Benenson et al., 2008, Levy et al., 2013] drivers are assumed to have some knowledge about the parking search area. Until finding a vacant parking spot, they cruise between temporary targets that are randomly generated within the destination's neighborhood, choosing, at every junction, a link whose other end is the closest to the next target. Levy et al., [2013] also assume that the decision to park at a vacant spot depends on the distance to destination and the level of parking occupancy observed over several previously traversed links – the higher the observed occupancy, the higher the probability to park further away from the destination. Mendoza-Silva et al. [2019] developed a simulation model to aid the design of a smart parking reservation system in a university campus. The system recognizes and reserves the vacant parking spot that is closest to a driver's destination and directs 'guided' agents there along the shortest path, whereas 'explorer' agents follow the PARKAGENT search tactic with target destinations as campus buildings.

We are not aware of a quantitative description of drivers' cruising behavior that is based on field observations and therefore rely on the outcomes of a serious parking game for this work.

### 2.4. Estimating drivers' parking behavior in serious games

Serious Games (SG) elicit human behavior through the immersion of players into virtual dynamic choice environments where players engage and interact in a carefully designed and fully controlled artificial conflict, and generate a quantifiable outcome [Michael and Chen 2005; Salen and Zimmerman 2004]. Players gradually learn how to behave in the game and gain experience in a way that replicates real-world learning. Game behavior can be consistently observed and, after being analyzed and formalized, interpreted as behavioral rules of agents in simulation models [Bonsall and Palmer 2004]. SGs can thus be considered a middle-ground between field observations that are difficult to manipulate, control and assess, and stated choice questionnaires which lack dynamism and realism.

In respect to the parking search, Ben-Elia et al. [2015] have recently proposed PARKGAME, a flexible SG platform for studying drivers' cruising behavior and decision-making in a virtual urban road network. In the game, a player drives the car to the destination and tries to park on-street or at the nearby parking facility. The road network is presented by arbitrary GIS layers, guaranteeing fast substitution of the experimental areas. Parking prices are presented to the player via the game's user interface. The player navigates using the arrow keys and sees up to 5 parking spaces ahead, occupied or vacant, and can park if the car's speed is below 12kph, similar to real driving. Our model of parking search behavior is based on the analysis of the PARKGAME outcomes.

### 2.5. Empirical observations of parking choice and behavior

Empirical studies of drivers' parking behavior are mostly based on stated preferences and to a lesser extent on revealed preferences. These studies aim at revealing drivers' preferences in regards to the parking price, location, type and distance from destination [van der Goot, 1982; Axhausen and Polak, 1991; Hess and Polak, 2004; Hunt and Teply, 1993; Lambe, 1996; Ma et al., 2013]; the potential impact of parking-restriction policies on the mode of travel and trip cancellation [Hensher and King, 2001; van der Waerden et al., 2006, Simićević et al., 2013; Shiftan and Burd-Eden, 2001; Shiftan and Golani, 2005; yan et al., 2019]; and the impact of information provision on parking behavior [Axhausen et al., 1994; Bonsall and Palmer, 2004; Lee et al., 2017; Chaniotakis and Pel, 2015].

Parking search time becomes essentially non-zero only in case of high occupation rates. This is clearly demonstrated by Van Ommeren et al. (2012) who averaged cruising time based on the Dutch National Travel Survey that combines different cities and parts of cities, and obtained the estimate of 36 seconds only. The authors explain this value by the identical pricing of on- and off-street parking in Dutch cities, and the low occupation rate may be the direct consequence of this pricing policy. Hampshire et al. (2016) report on the cruising behavior of 109 drivers in downtown Ann Arbor and downtown Detroit, Michigan. All drivers' cars were equipped with the GPS and videos cameras and the start of cruising was identified by the drivers' body language and driving speed. The paper does not report on the parking occupation rate in the vicinity of drivers' destinations, while one can assume that in downtown Ann Arbor and Detroit the on-street occupation rate is usually below links' parking capacity. Low occupation may explain why search time for 70% of the drivers was short, while the remaining 30% were responsible for 70% of the total cruising time. Despite lack of the data on the parking occupation rate, their paper highlights the role of individual variability in the cognition of search time and willingness to pay for parking.

As can be seen, simulation and theoretical models, even with the rough and heuristic estimates of drivers' cruising behavior, are quite successful in reproducing three major parameters of curb parking – occupation rate, distance between parking place and destination, and cruising time. However, generating these estimates for a real city demands numerous simulations of parking search for thousands of drivers and is extremely time consuming, especially when the influence of the model parameters should also be studied. The goal of this paper is to propose a simple approximate method for estimating one of these three major parameters of curb parking – parking search time.

### 3. PARKGRID, an Agent-Based Model of Parking Search

PARKGRID is an agent-based model that continues the tradition of PARKAGENT [Levy et al., 2013; Levy and Benenson, 2015]. It is a GIS-based application that builds on the layers of streets, destinations, and on-street parking spots and off-street parking facilities, and can be freely downloaded from https://www.researchgame.net/profile/Nir_Fulman. PARKGRID, like all agent-based models, is based on the rules of agents' behavior, in this case the drivers' parking search rules. These rules are based on experiments with an SG for parking search [Benenson et al., 2015]. In this paper, we apply PARKGRID to estimate parking search time in order to validate our algorithm.

### 3.1. Urban Street Network in PARKGRID

PARKGRID is based on the standard GIS representation of an urban road network, accounting for the traffic directions and parking restrictions on each street link. Optionally, a layer of destinations, represented by building centroids, and a layer of parking facilities can be used.

### 3.2. Abstract grid city in PARKGRID

We start with investigating parking search in an abstract grid city of two-way streets (Figure 1). The length of a street link is 100m. To avoid boundary effects, the grid is folded into a torus - the right ends of its rightmost links in Figure 1a are connected to the leftmost junctions and the top ends of the top links are connected to the junctions at the bottom. In this way each junction has exactly four incident links.

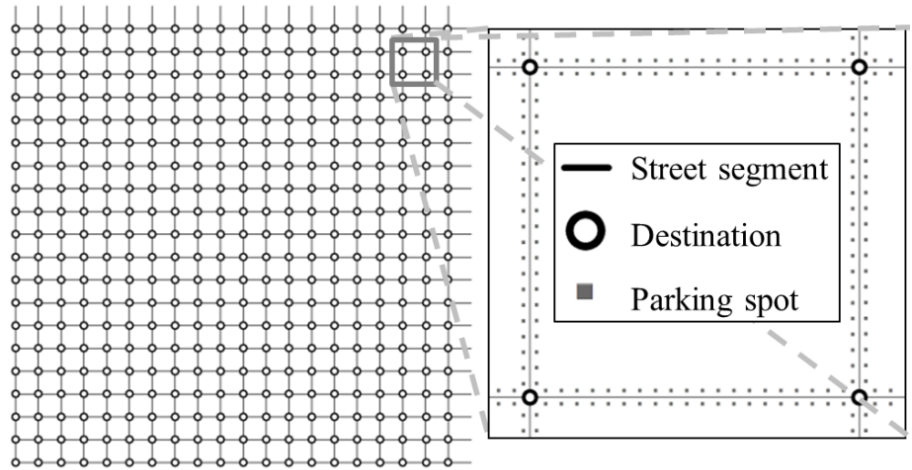

Figure 1: (a) Torus 20x20 grid city; (b) Zoom to a city block.

For further simplicity, we set drivers' destinations at the junctions. Each destination n is characterized by its demand d, which can vary between destinations. Each link l of the grid city contains 20 parking places of 5m length on each of its sides, 40 parking spots altogether. This entails the ratio R of the total number of destinations to the total number of curb parking spots equal to $R_{city}$ = 80.

Street links and junctions are stored as GIS layers, with the demand being an attribute of a junction, and the number of parking spots an attribute of a link. Model experiments are performed on a 20x20 grid with N = 400 destinations (junctions) and L = 800 links. The number of parking places P in the grid city is thus equal to L*40 = 32,000. In this paper, the PARKGRID model does not include off-street parking facilities.

### 3.3. Drivers as agents

PARKGRID agents (drivers) are explicitly considered from the moment they are close to their destination and start cruising for parking; drivers enroute are ignored. While cruising, a driver in the model either finds a vacant parking spot and parks, or leaves the system after a long unsuccessful search.

We assume that a driver cruises at a constant speed of 12 km/h [Carrese et al., 2004] that is, it takes a driver 30 seconds to traverse a 100m link. We thus consider a 30 second interval as the model's time unit - tick. At each tick, the list of cruising and due-to-depart drivers is constructed, randomly re-ordered, and each driver acts according to its place in the list, facing the parking pattern created by the actions of its predecessors.

### 3.4. Drivers' cruising behavior as a biased random walk

As discussed in section 2, not much is known about drivers' cruising behavior. To establish driver's behavior for PARKGRID agents, we rely on the outcomes of the PARKGAME SG [Benenson et al., 2015]. A player of PARKGAME uses the keyboard to drive the car i.e., advance it in space, accelerate, decelerate, turn, stop and park. The game's goal is to arrive at a predefined destination building within 3 minutes from beginning the search. Players receive a prize for each successful game, but are fined for each late arrival. Additional rewards are given when players find free on-street parking over more expensive (but possibly closer) off-street parking. An external observer will notice that game players initially loop around the destination, repeatedly approaching it and driving further away until finding a free on-street parking spot (Figure 2a) or stop cruising and pay for parking in an off-street facility The latter may happen at the very end of a game, when the player must choose between being penalized for a late arrival, paying for off-street parking, or parking illegally and incurring a parking fine that is significantly higher than the parking fee. It's important to note that although U-turns were allowed in the game, players never took them, therefore we ignore this option.

In [Fulman et al., forthcoming 2019] we demonstrate that players' behavior in PARKGAME can be formalized in terms of a biased, towards destination, random walk [O'Sullivan and Perry, 2013]. To define this kind of behavior, let us consider a player/driver who fails to park on the link just traversed, reaches the next junction and must decide whether to turn left, right or continue straight. First, as for the standard biased towards destination random walk model, a driver's turn decision depends on the distance to the destination. For longer distances, the probability to choose the link that will take the driver closer to the destination increases, while the probability to drive further away decreases. Second, and different from the standard biased random walk model, a driver's turn decision on reaching a junction depends on the turn decision taken at the previous junction: A driver who is approaching the destination will more probably choose the link that leads closer to the destination, while the driver who is moving away from the destination will likely continue outbound rather than inbound. The probabilities to turn closer to/further away from the destination, as dependent on the distance to destination and the decision made at the previous junction, are based on 230 PARKGAME games, played by 29 participants (Table 1). These probabilities were estimated for several levels of parking occupation, all above 95%, and were not dependent on the parameters of the game. Note that as a result of the above behavior, drivers always remain within a neighborhood $U(n)$ of 500m radius around their destinations $n$ (Table 1).

Table 1: Probability that a PARKGAME driver cruising for parking who failed to park on the link just traversed, will choose the next link that leads closer to/further away from the destination, as dependent on the distance d between the current junction and the destination and the decision made at the previous junction [Fulman et al., forthcoming 2019].

| Decision at the previous junction | d = 0 | | d = 100 | | d = 200 | | d = 300 | | d = 400 | | d = 500 | |
|---|---|---|---|---|---|---|---|---|---|---|---|---|
| | Closer | Further | Closer | Further | Closer | Further | Closer | Further | Closer | Further | Closer | Further |
| Closer | Irrel | 1.00 | 0.65 | 0.35 | 0.85 | 0.15 | 0.90 | 0.10 | 0.90 | 0.10 | Irrelevant | |
| Further | Irrelevant | | 0.00* | 1.00 | 0.80 | 0.20 | 0.85 | 0.15 | 0.85 | 0.15 | 1.00 | 0.00 |

*Driving closer to the destination from the junction at a 100m distance demands a U-turn. This option was never chosen by the players

The probability to drive closer to/further away from the destination, as presented in Table 1, determines the probabilities $p_c(l)$ of traversing links $l \in U(n)$ as dependent on the minimum distance d between a link l and destination n (Figure 2b). Figure 2c presents the fraction of cruising time that a driver spends at the distances [0, 100), [100, 200), … from the destination, and the approximation that is chosen to preserve the maximal value of 0.58 for the probability to cruise at distances between 0 and 100 m.

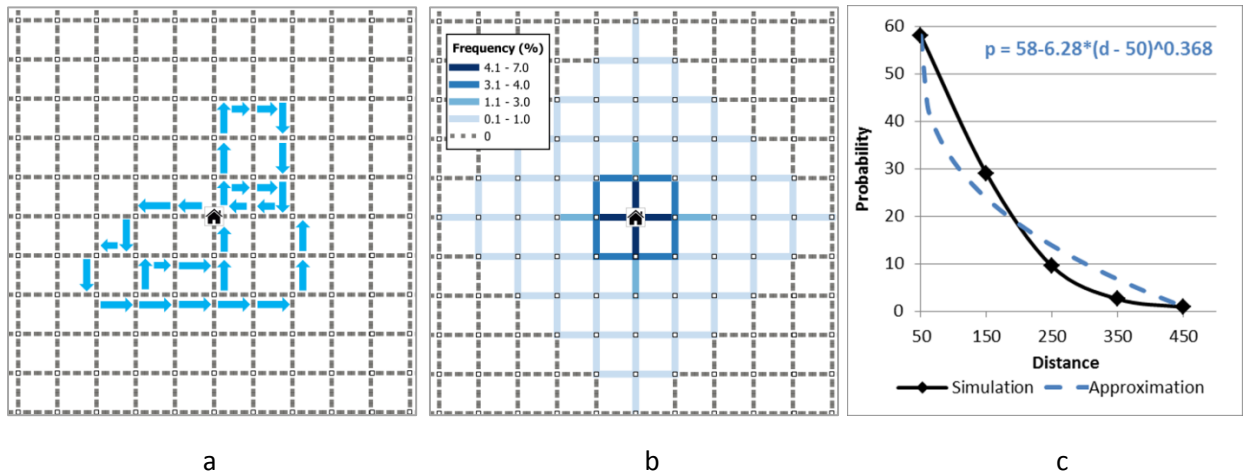

a            b            c

Figure 2: (a) PARKGAME player's search paths; (b) Probabilities $p_c(l)$ of traversing links $l \in U(n)$ while cruising; (c) The fraction of time drivers cruise at a certain distance from the destination.

The results of PARKGAME were incorporated into the PARKGRID model to reflect human drivers' cruising behavior. At every 30-seconds tick t, a model agent (driver) that fails to park on the link traversed between t -1 and t, reaches the next junction and decides on the next link to cruise. The choice of link follows the biased random walk model described above. If k drivers $c_i$ have chosen the same link l for traversing, the order $c_1, c_2, … c_k$, in which they traverse l is established randomly, and the number of vacant parking places on l is updated during the tick. In the case where link l has $f_l(t)$ vacant parking places at t, then driver $c_1$ will park there and its search time on this link is estimated as $30/(f_l(t) + 1)$ seconds. Driver $c_2$ is then considered as traversing l with $f_l(t) - 1$ vacant parking places, etc. After l becomes fully occupied by the first $f_l(t)$ drivers, the remaining $k - f_l(t)$ drivers fail to park on that link, then reach the next junction, and each driver choses its next link as described in the behavioral model above.

### 3.5. Drivers' arrivals and departures

In the PARKGRID model we consider a city of "office buildings" and the drivers who aim to park in the area are divided into *employees* and *visitors*. Employees arrive to the city in the morning and do not leave until the end of the simulation; visitors arrive to the city for 1-2 hours. Destination's $n_i$ hourly demand for parking $d_i$ is defined as the total number of employees who aim at $n_i$, plus the average hourly number of visitors who also aim at $n_i$. Below we assume that for each $n_i$, employees comprise a constant fraction e of the demand $d_i$. To recall, the total parking supply in the city is constant and equal to L*40 = 32,000 spots; we thus investigate drivers' cruising time as dependent on the total demand $D = \sum\{d_i, i = 1, \ldots 400\}$ and its spatial pattern $D_S$, defined by the demand at each junction $n_i$. The average occupation rate q in the city is q = D/L.

Simulation of the working day starts at 8:00 with an empty city and stops at 15:00. Employees arrive to the city in the morning between 8:00 and 9:00, and park until the end of the day. Visitors arrive and depart throughout the entire day, and their parking time is uniformly distributed on the [$\tau_{min}$, $\tau_{max}$] interval, where $\tau_{min}$ = 1 hour and $\tau_{max}$ = 2 hours. Visitors vacate their spot at the end of their parking time. Note that the assumption of uniform distribution of the visitors' parking time is different from the standard view of a constant departure rate [e.g. Caicedo et al., 2012, Levy et al., 2015], and entails a piecewise linear and not exponentially decreasing probability of a parking place being occupied for longer than $\tau$ ticks.

PARKGRID model drivers aiming at destination $n_i$ arrive to the city according to a Poisson process with a per-hour average $\lambda_i$ that is proportional to the $n_i$'s demand for the driver's category (employee/visitor). On arrival, driver c starts its parking search at the beginning of one of the links l within the $U(n_i)$. The starting link is chosen according to the probability $w_l$ of visiting this link during the parking search (Figure 2b). The driver starts cruising for parking according to the rules presented in section 3.4. Maximum cruising time M is set to M = 20 minutes in all investigated scenarios; during this time a driver traverses 40 links and 1,600 parking places. The drivers who fail to find curb parking during time M, park at a "distant off-street parking facility" that is never full. We ignore these drivers when estimating the average cruising time.

Given the total demand D and its spatial pattern $D_S$, we generate a city with the given average occupation rate q by setting the employees' arrival rate to between 8:00 and 9:00 for each destination $n_i$ as $\varepsilon_i = e*q*d_i$, e < 1, and then adjust the arrival rate of the visitors to $n_i$, in order to guarantee that the average number of visitors arriving to $n_i$ and searching for nearby parking is equal to $(1 - e)*q*d_i$ per hour. Accounting for the average visitors' parking time $(\tau_{min} + \tau_{max})/2$, the per hour arrival rate of visitors to destination $n_i$ should be thus $\lambda_i = 2*(1 - e)*q*d_i/(\tau_{min} + \tau_{max})$.

No matter what is the spatial pattern $D_S$ of the demand, the average occupation rate q(t) in the city converges, towards the midday, to the value fluctuating around the preset q = D/L with very low variation. In what follows, we investigate drivers' cruising time as dependent on q and on the spatial pattern of the demand.

Our goal in this paper is to estimate the "cruising time curve" - the probability $p(\tau, n)$ for a driver that aims at a destination n to search for parking longer than time $\tau$, as dependent on the demand and supply patterns around n. We start our analysis with the case of a uniform distribution of demand and supply in the city, and then extend these results to the case of heterogeneous demand and homogeneous supply. The case of heterogeneous demand and supply is similar to the latter one. We repeat every computational experiment 100 times and present the average of the parameters obtained in these experiments. The variation in results over these experiments is very low (CV < 0.5%)

## 4. Cruising Time Curve for the Grid City

### 4.1. Cruising time curve for a city of homogeneous demand and supply

In this section we consider a grid city with an equal demand for each destination $n_i$, $d_i = q*R_{city}$. The average occupation rate throughout the city stabilizes at q towards 10:00, an hour after the employees arrive, and remains steady from then on, fluctuating around q with a very low STD of ~0.0025 that is not dependent on q. In what follows, we establish the cruising time curve $p_q(\tau, n)$ for the drivers that search for parking during the period of time between 10:00 and 15:00, when the average occupation rate is steady and the occupation pattern is defined by the outcomes of drivers' arrivals, departures and searches. We start with the employee:visitor ratio 85:15 that is, e = 0.85, and then investigate the sensitivity of results using e = 0.55 and e = 0.25.

For a low average occupation rate q < 0.75, the occupation rate $q_l(t)$ of each link l remains below 1 almost all the time, and the average fraction of fully occupied links for which $q_l(t) = 1$ is lower than 0.1%. In these circumstances, almost every driver finds a vacant parking spot on the link where the search begins, and the overall number of cruising drivers is very low. With an increase in q, the average per link number of simultaneously cruising drivers grows non-linearly (Figure 3a) and the parking pattern qualitatively changes. Note that the fraction of fully occupied links grows significantly faster than can be expected based on the random distribution of the same number of cruising drivers over the links of the grid city (Figure 3b).

The fully occupied links of the parking pattern are clustered into patches, a characteristic of parking search patterns. The emergence of clusters of fully occupied links and, in parallel, clusters of relatively sparsely occupied links was initially recognized by [Levy et al, 2013] and is inherently related to the parking process. Namely, a driver that during [t, t + 1] has traversed the fully occupied link $l_1$, and then enters at t + 1 to the link $l_2$ that has a vacant spot, will park at $l_2$. This phenomenon strengthens with the increase in q and, as a result, the number of fully occupied links of the parking pattern is higher than that for the scenario of a random assignment of cars to spots. We characterize this clustering by the correlation between a link's occupation rate and the average occupation rate of the links navigable from it (Figure 3c). Figure 4 presents the emerging clusters for q = 0.95 in comparison with the parking pattern for q = 0.85.

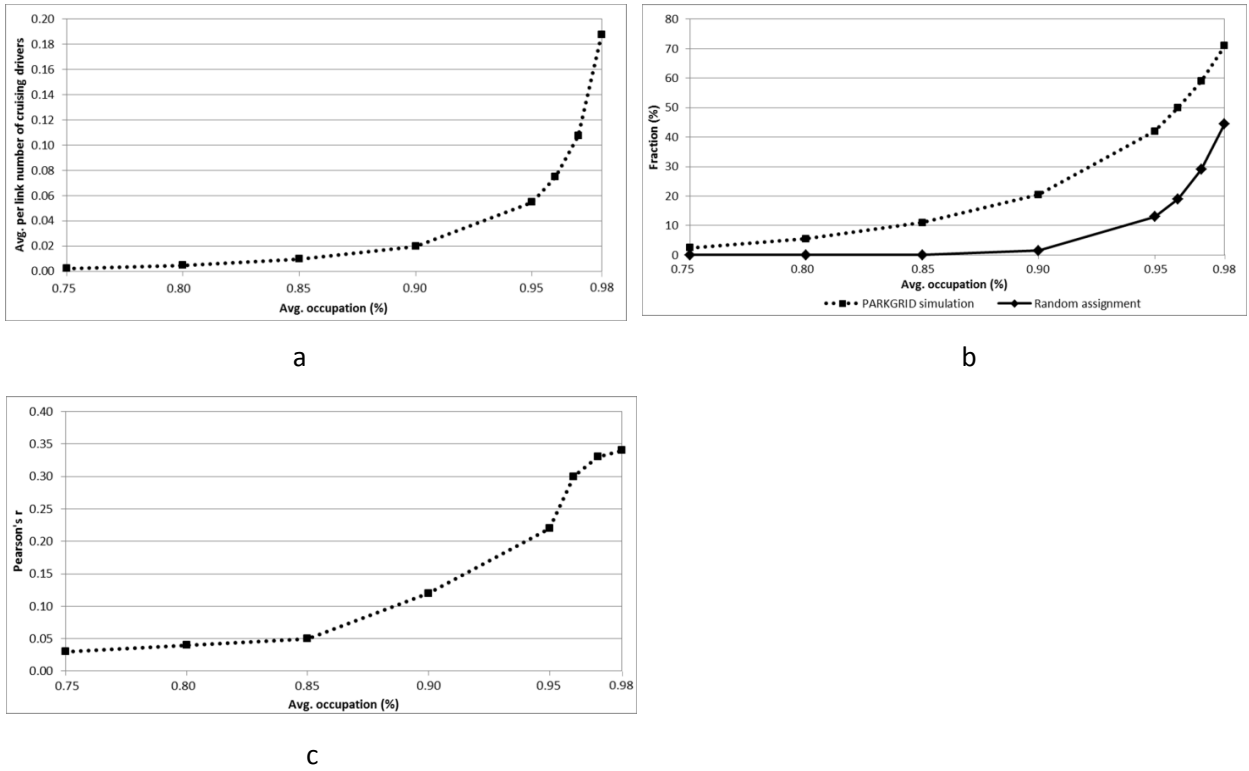

Figure 3: Grid city parking pattern parameters at t = 15:00: (a) Average per link number of cruising drivers in the grid city; (b) average fraction of fully occupied links in the simulation and for the random selection of a parking spot; (c) correlation between occupation rate of a link and the average of occupation rates of the links navigable from it, as dependent on q

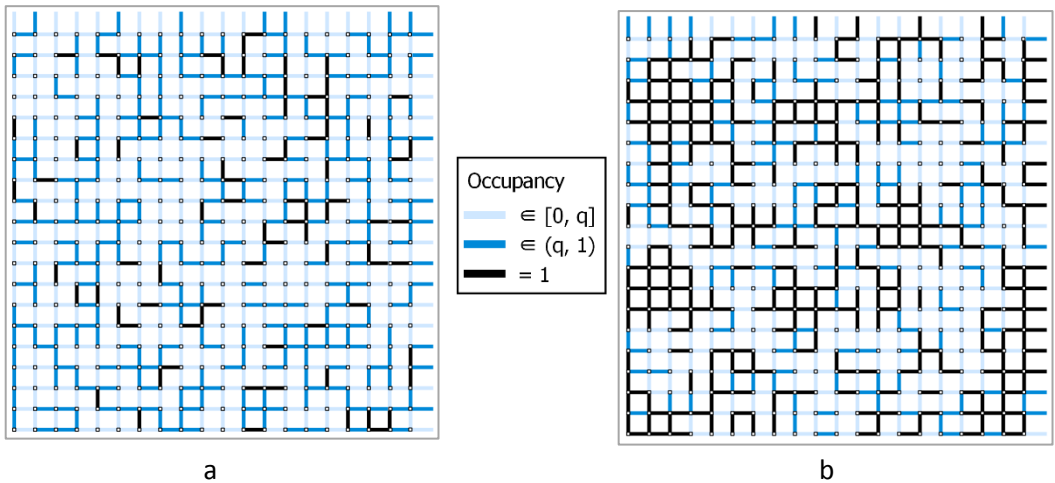

Figure 4. The PARKGRID patterns of fully occupied links for (a) q = 0.85 and (b) q = 0.95

Clusters emerge and dissolve in time, but for realistic arrival and departure rates, their shape changes slowly. This is confirmed by the linear autocorrelation function between link's occupation at t and t + k for k ≤40 (20 minutes, the maximal duration of a parking search) that decreases very slowly (Figure 5).

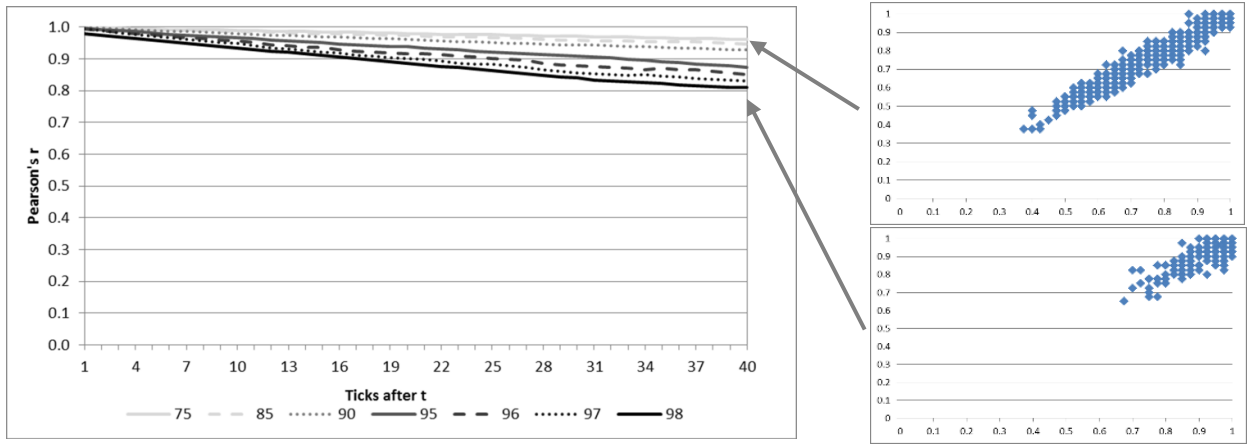

Figure 5. Autocorrelation function for a link's occupancy and a scatterplot of the link's occupancy at tick = 40 vs occupancy at tick = 1. Top q = 0.75 (r = 0.96); bottom q = 0.97 (r = 0.83)

Since a driver cannot park on a fully occupied link, we hypothesize that the number of fully occupied links within the destination's search neighborhood can determine the parking search time there. To verify this hypothesis, we introduce the *wasted search time index* $F_{U(n)}$ that is an average fraction of the search time that a driver whose destination is n would spend traversing fully occupied links within the U(n). $F_{U(n)}$ is just a sum of probabilities to visit fully occupied links within the U(n):

$$F_{U(n)}(t) = \sum_{l \in U(n)} \{p_c(l) | f_l(t) = 0\} \qquad (1)$$

Over all destinations of the homogeneous city, the average value of $F_{U(n)}$ grows with the increase in q just as the fraction of fully occupied links grows (Figure 3a). The same is true for the temporal autocorrelation between the values of $F_{U(n)}(t)$ and $F_{U(n)}(t + k)$ (Figure 5).

To verify whether $F_{U(n)}$ determines the cruising time curve p(τ, n), let us estimate the conditional probability of parking failure $p_{failure}(F)$ within the neighborhood U(n) for which $F_{U(n)}$ remains almost constant, $F_{U(n)} \in [F − 0.01, F + 0.01]$ during several ticks, from the moment a driver's parking search starts (Table 2). As can be seen, $p_{failure}(F)$ remains constant in time, that is, it does not depend on the spatial pattern of fully occupied links within U(n). In addition, and as can be expected, $p_{failure}(F)$ is very close to F (Table 2).

Table 2. Conditional probability $p_{failure}(F)$ of parking failure at t for drivers searching for parking within the neighborhood U(n) for which $F_{U(n)}$ remains constant from the start of the parking search until t

| t \ $F_{U(n)}$ | 1 | 2 | 3 | 4 | 5 | 6 | 7 | 8 | 9 | 10 | Average | STD |
|---|---|---|---|---|---|---|---|---|---|---|---|---|
| **0.05** | 0.05 | 0.06 | 0.07 | 0.05 | 0.04 | 0.04 | 0.05 | 0.06 | 0.06 | 0.05 | **0.053** | 0.0095 |
| **0.10** | 0.09 | 0.10 | 0.09 | 0.10 | 0.11 | 0.11 | 0.11 | 0.12 | 0.10 | 0.11 | **0.104** | 0.0097 |
| **0.20** | 0.20 | 0.20 | 0.21 | 0.20 | 0.21 | 0.21 | 0.20 | 0.20 | 0.20 | 0.21 | **0.204** | 0.0052 |
| **0.40** | 0.39 | 0.42 | 0.39 | 0.41 | 0.41 | 0.40 | 0.42 | 0.39 | 0.41 | 0.40 | **0.404** | 0.0117 |
| **0.60** | 0.59 | 0.60 | 0.61 | 0.62 | 0.60 | 0.61 | 0.60 | 0.60 | 0.62 | 0.62 | **0.607** | 0.0106 |
| **0.80** | 0.81 | 0.82 | 0.80 | 0.79 | 0.79 | 0.81 | 0.82 | 0.80 | 0.81 | 0.81 | **0.806** | 0.0107 |
| **0.90** | 0.90 | 0.90 | 0.91 | 0.89 | 0.89 | 0.89 | 0.91 | 0.90 | 0.90 | 0.89 | **0.898** | 0.0079 |
| **0.95** | 0.96 | 0.95 | 0.95 | 0.94 | 0.93 | 0.95 | 0.96 | 0.95 | 0.94 | 0.94 | **0.947** | 0.0095 |

For a driver cruising within U(n), for which $F_{U(n)} \approx F$, cruising time decays geometrically:

$$p(\tau, n) = (1 - F) * F^{\tau} \qquad (2)$$

Given q, $F_{U(n)}$ is distributed as presented in Figure 6, and can serve as a basis for estimating the cruising time curve $p_q(\tau)$ for the entire city. Namely, for the average occupation rate q

$$p_q(\tau) = \sum_{F \in (0, 1)} \{(1 - F) * F^{\tau-1} * w_q(F)\} \qquad (3)$$

where $w_q(F)$ is the frequency of search neighborhoods with the $F_{U(n)} = F$. Practically, we apply (3) for the intervals of F of the 0.02 width.

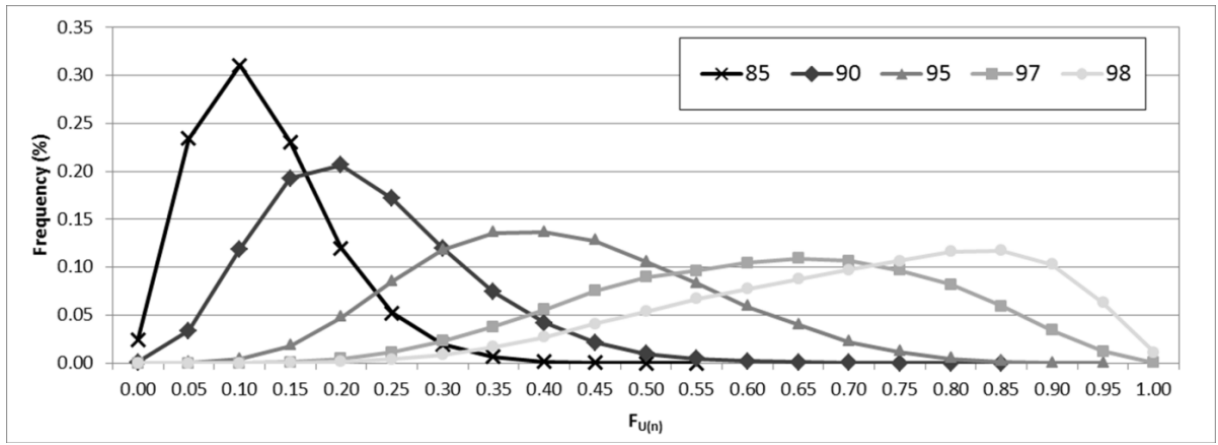

Figure 6. Distribution of $F_{U(n)}$ as dependent on q

The comparison between the cruising time curves estimated according to (3) and as generated by the PARKGRID simulations demonstrates a very good fit (Figure 7).

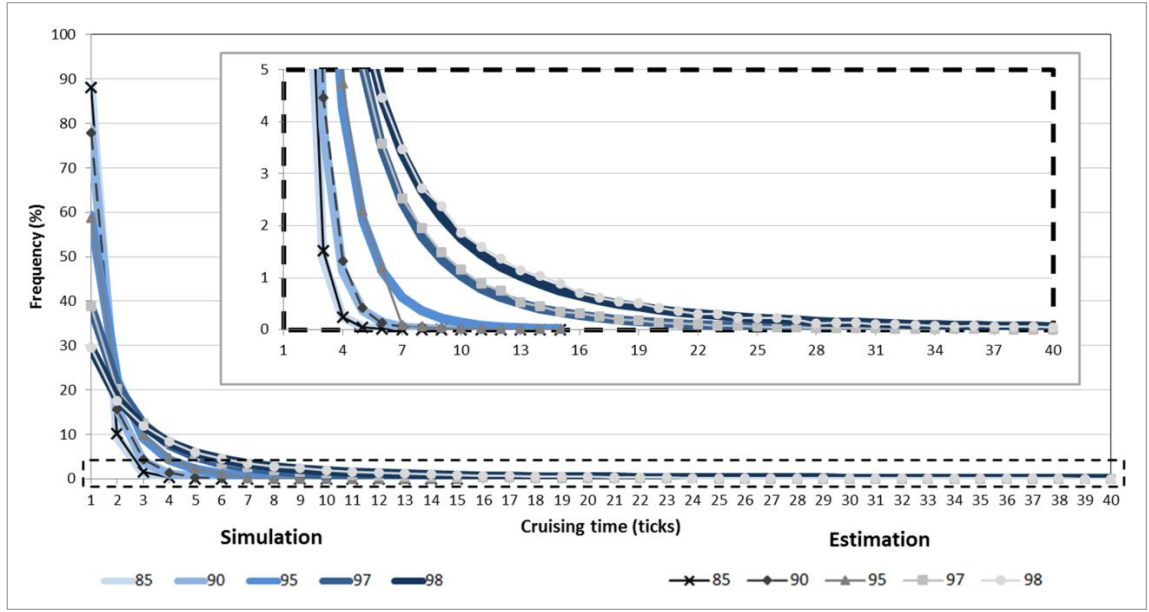

Figure 7: Cruising time curves for the grid city as a whole, for q varying between 0.85 – 0.98 and e = 0.85

### 4.2. The Case of Heterogeneous Demand and Homogeneous Supply

Formula (2) defines the cruising time curve for the driver whose destination is n as dependent on the $F_{U(n)}$ only. That is, it can be directly applied to the case of heterogeneous demand. To test this, we consider a city in which the average demand-to-supply ratio over the city remains equal to q, except for two neighborhoods H and L, where the demand-to-supply ratio differs from q. In neighborhood H, the demand $d_i$ of each destination is higher than $q*R_{city}$, while in neighborhood L, the demand $d_i$ is lower, to compensate for the increase in demand in H. Formally, for each destination $n_i \in H$ the demand $d_i = (q + α)*R_{city}$, while for $n_i \in L$, $d_i = (q − α)*R_{city}$.

Figure 8a presents the case of 5x5 H and L neighborhoods, for q = 0.85 and α = 0. 5. The demand of destinations in H is equal to $d_i = (q + α)*R_{city} = (0.85 + 0.5)*80 = 108$, while the demand for destinations in L is $d_i = (q − α)*R_{city} = (0.85 − 0.25)*80 = 28$. For the rest of the destinations $d_i = q*R_{city} = 0.85*80 = 68$.

The effects of the H and L neighborhoods on the city parking pattern are different. The L neighborhood hardly influences the parking pattern, because drivers aiming to park there, easily park on one of the links adjacent to the destination. However, the insufficient parking capacity of the links within H forces some drivers to park beyond the neighborhood, increasing parking occupation in H's surroundings (Figure 8b).

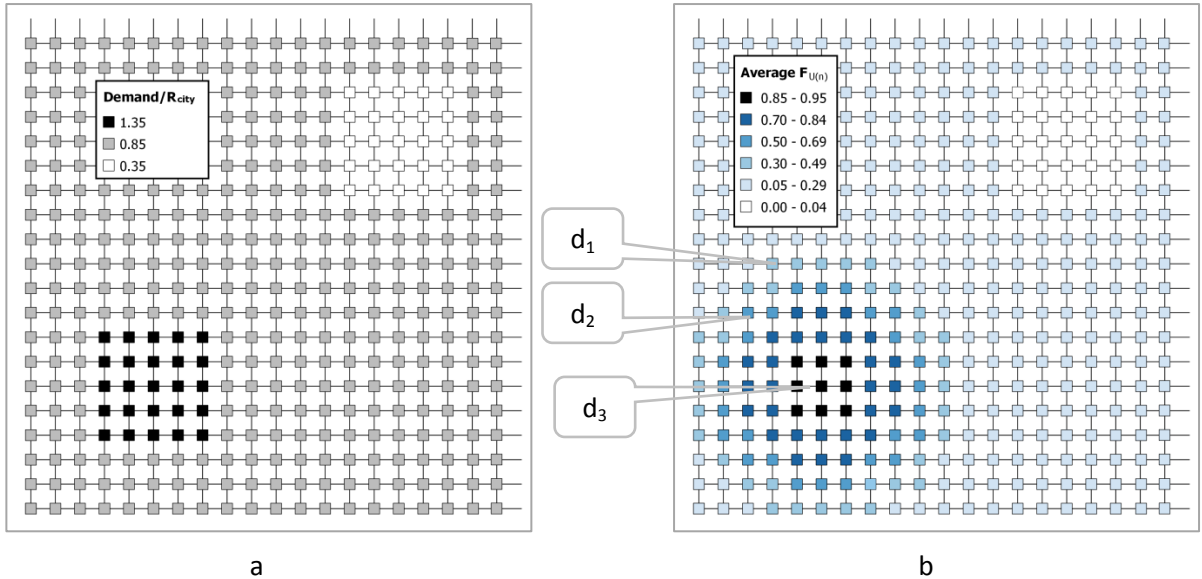

Figure 8: (a) Heterogeneous demand scenario, q = 0.85, e = 0.85, R = 80, α = 0.25; (b) The values of $F_{U(n)}$ for the parking pattern for a steady average occupation rate, as generated by the PARKGRID. The three destinations $d_1$, $d_2$, $d_3$ are exploited in Figure 9.

It is important to note that the links within the H neighborhood and its nearest surroundings are almost always fully occupied, while the patches of 100% occupation over the rest of the city grow larger with the increase in q, and emerge and dissolve in time, just as in the homogeneous case.

Figure 9 presents the cruising time curves for those destinations within the H spillover that are characterized by the same $F_{U(n)}$, as generated by the PARKGRID, and the approximations of these curves obtained with (2). As can be seen, they are very close.

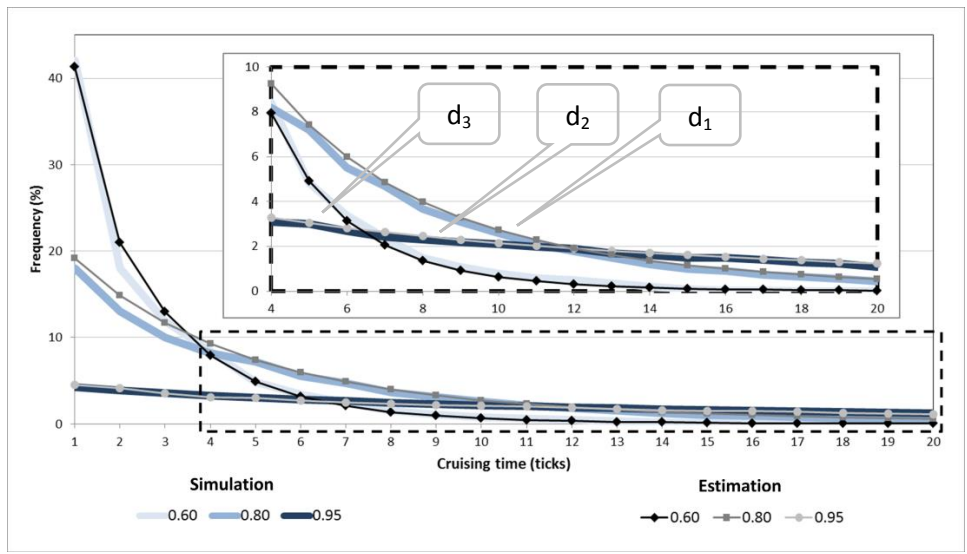

Figure 9: Cruising time curves estimated according to (2) and as generated by the PARKGRID, for three destinations with different $F_{U(n)}$ within H and its surroundings, for q = 0.85, 0.90, 0.95 and e = 0.85

To estimate parking search time for the driver whose destination is n, one has to know the $F_{U(n)}$ value. In the case of random demand and supply, constructing the occupation pattern demands multiple simulation runs. We therefore consider a simplified occupation pattern that combines both demand and supply and is sufficient for approximating p(τ, n). This *Maximally Dense (MD)* parking occupation pattern can be generated with the extended version of the PARKFIT algorithm proposed by Levy and Benenson [2015].

### 4.3. Approximation of Parking Occupation by Maximally Dense Parking Pattern

The MD pattern generating algorithm considers a city of "connected automated cars," each of which, instead of cruising for parking, follows the direction of a central dispatcher that, at the moment the car claims the necessity to park, assigns the closest vacant parking spot to the car's destination. Once assigned, this spot cannot be occupied by other cars.

Formally, at each tick t, the algorithm consists of four steps:

(1) Generate a list $G_t$ of <car, destination> pairs for all drivers that arrive at t and randomly reorder it. $G_t$ defines the order of cars' arrivals at t.
(2) Loop by cars in $G_t$. For each car, consider the vacant parking spot that is closest to its destination and park there.
(3) Release spots that were occupied by cars whose parking time is over at t.
(4) Repeat stages (1) – (3) until the overall occupation pattern stabilizes

The occupied spots in the obtained pattern are concentrated around the destinations, and we thus call it the Maximally Dense (MD) pattern.

### 4.4. Cruising Time Curve Based on the Maximally Dense Pattern

Given an average occupation rate q in the city, the distribution of $F_{U(n)}$ for the MD pattern is very close to that obtained in the PARKGRID simulations (Figure 10) and we can exploit this for estimating $w_q(F)$ in (3).

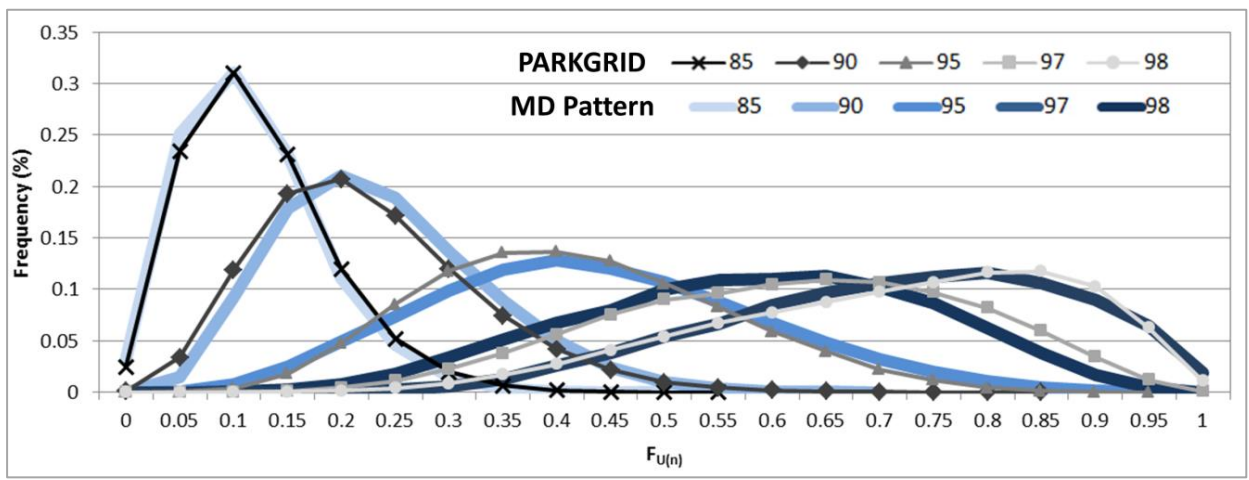

Figure 10: Distributions of $F_{U(n)}$ for the pattern of parking demand in Figure 4, generated by the PARKGRID and based on the MD pattern.

Figure 11 presents the comparison between the cruising time curves generated by the PARKGRID and approximate ones for all 400 destinations $n_i$ of the two-neighborhood city in Figure 8. The estimates of the average cruising time and of the probability to cruise longer than 3 minutes also strongly correlate, with $R^2 \sim 0.95$ in both cases.

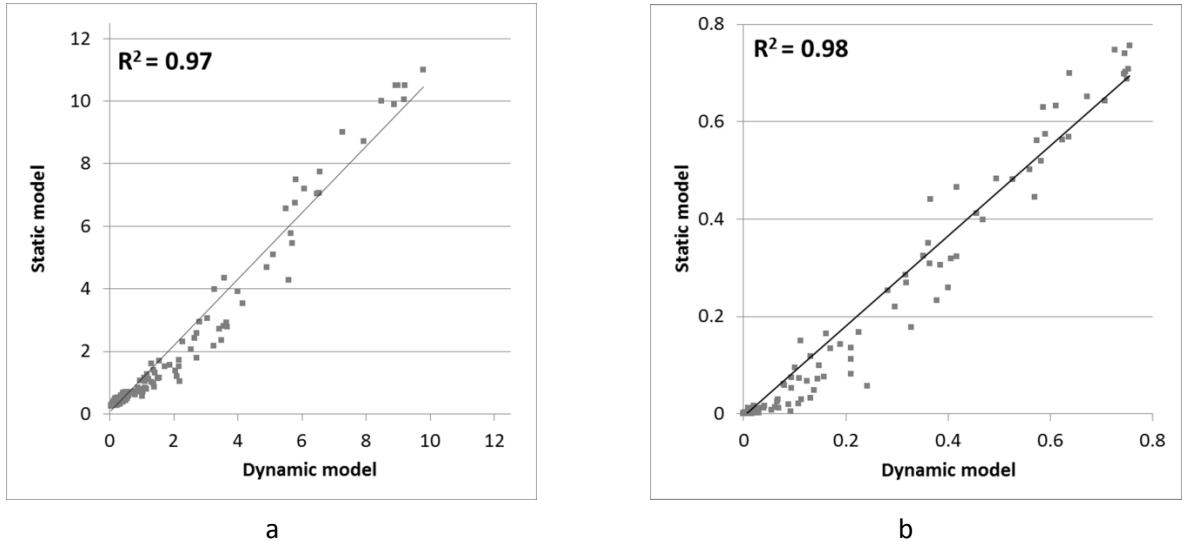

a                                                                                       b

Figure 11: Comparison between cruising time curves generated by the PARKGRID and approximate ones, for destinations in the heterogeneous city (a) Average search times (minutes); (b) Probability to cruise longer than 3 minutes.

The MD pattern generating algorithm enables estimating cruising time curves for all destinations in large urban areas. Note that as by-products, we can obtain distribution of the distance to parking and the parking occupation pattern. In what follows we focus on the cruising time only, and apply the proposed approach for estimating evening curb parking search time in the Israeli city of Bat Yam.

## 5. Predicting Cruising Time in Bat Yam

As a practical example, we estimate the search time of residents' search for overnight parking in the Israeli city of Bat Yam. City population data as well as layers of streets, off-street parking facilities and buildings were supplied to us by the Bat Yam municipality.

### 5.1. Parking Demand and Supply in Bat Yam

Our estimates are based on the demand and supply data of 2010, when Bat Yam's population was ca. 130,000, total car ownership 35,000, and the total number of residential buildings 3,300 with 51,000 apartments. Residential buildings in Bat Yam provide their tenants a total of 17,500 dedicated off-street parking spots that should be excluded from the demand and supply data.

Parking supply data is based on two GIS layers - a layer of streets and a layer of off-street parking facilities. Based on the layer of streets, 27,000 spots for curb parking were constructed automatically, 5 meters apart on both sides of two-way street links, and on the right side of one-way links, with a necessary gap from any junctions. In addition, 1,500 spots are available for the city's residents in its parking facilities, where Bat Yam residents can park in the evening free of charge. The average overnight demand/supply ratio is thus very low (35,000 − 17,500) / (27,000 + 1,500) ≈ 0.61 car/parking spot.

However, the distribution of demand and supply in Bat Yam are both highly heterogeneous, and the demand in the center of Bat Yam significantly exceeds the supply there (Figure 12).

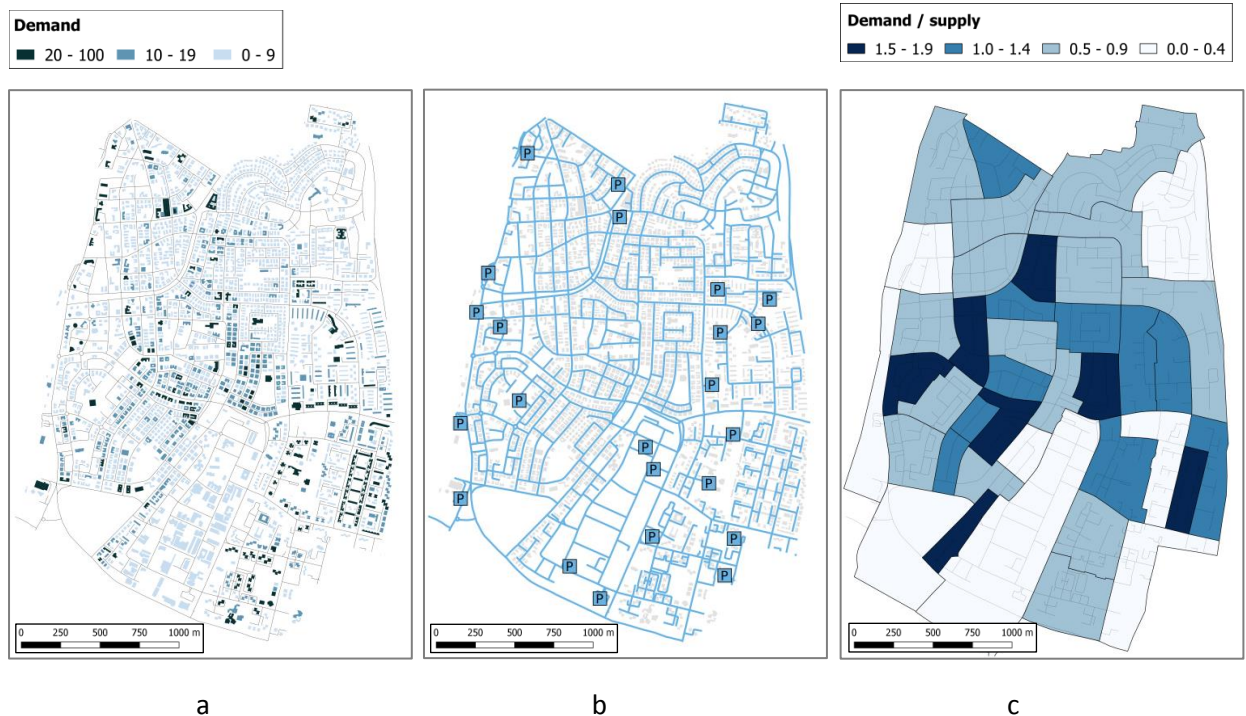

          a            b            c

Figure 12: Bat Yam: (a) Parking demand by buildings; (b) Parking supply represented by street links and off-street facilities; (c) Demand-to-Supply ratio aggregated by Transport Analysis Zones.

### 5.2. Maps of Parking Search Time in Bat Yam

We estimate cruising time in Bat-Yam in the evening, assuming the ratio between residents and visitors seeking parking is 90:10. This is done in four steps:

1. Estimate parking demand by buildings, and supply by urban links and facilities (Figure 12a)

2. Transfer buildings' demand to the junction that is nearest to the building

3. Construct Bat Yam MD pattern for the 90:10 resident:visitor ratio

4. For each destination building n, and each link l estimate the distance between l's midpoint and n. Consider links l, for which this distance is less than 500m.

5. For each destination n build an $F_{U(n)}$ pattern by applying approximated probability of cruising at a link l at a distance d from n as revealed in the PARKGAME experiments (Figure 2c).

6. Estimate cruising time curve for each destination n, as dependent on the $F_{U(n)}$, applying formula (3)

Figure 13 presents the patterns of the wasted search time index, the average search time in Bat Yam, and the probability to cruise for parking for more than a certain time for the city's northern part.

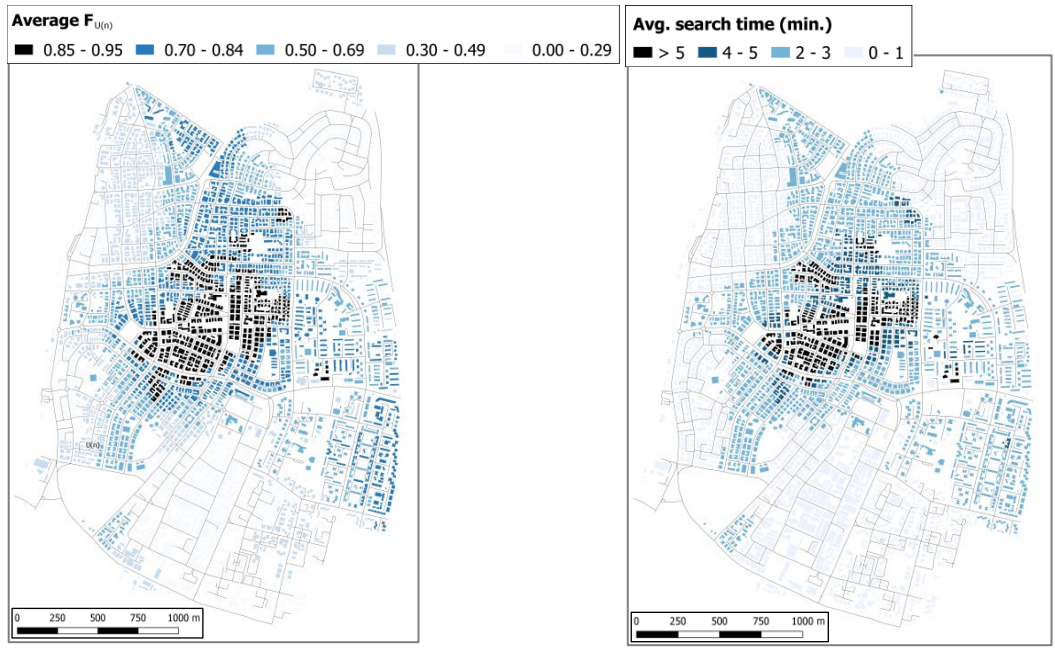
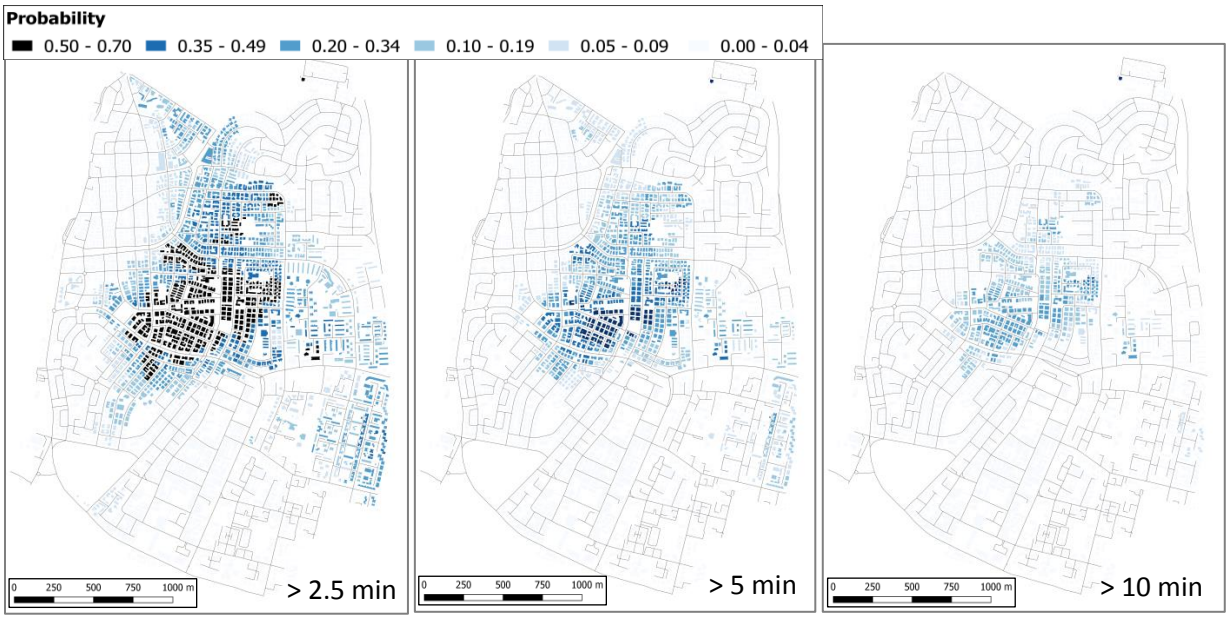

Figure 13. Bat-Yam: (a) The value of the wasted search time index $F_{U(n)}$; (b) average cruising time and the probability to cruise longer than (c) 2.5 min; (d) 5 min; and (e) 10 mins.

## 6. Discussion and conclusions

We propose an algorithm for the approximate estimation of a drivers' cruising time that is based on the high-resolution maps of a city's parking demand and supply, and the probability to search closer or further away from the destination estimated in a serious game. As a by-product, we obtain the distribution of the distance to destination and parking occupation. The proposed method can be applied to every city where the patterns of parking demand and supply are known at a resolution of buildings, roads and parking facilities. The estimates obtained are very close to those obtained in a dynamic simulation model. Applying our algorithm to the Israeli city of Bat Yam, we demonstrate that despite a low, ca. 61%, average demand to supply ratio, spatial heterogeneity of the demand and supply patterns results in lengthy parking searches for a significant fraction of drivers. It should be stressed that the perspectives of agent-based modeling of human-driven systems such as parking, critically depend on our knowledge of agents' behavior. In this respect, we consider serious games as a framework for capturing and further formalizing human behavior in dynamic self-organizing systems.

The paper does not include a study of the goodness-of-fit of the proposed approximation as dependent on the parameters of a parking system, such as the ratio of employees to visitors, the duration of parking, and variations in the rate of arrivals throughout the day. However, our preliminary investigation that focused on the sensitivity of the proposed approximation to the employee:visitor ratio clearly manifests low or very low sensitivity.

From a practical point of view, parking search time is a basic parameter for the urban planner who aims at assessing the consequences of construction of a new office, commercial or residential building. If parking supply in the area is insufficient for the anticipated demand, a planner can choose to increase supply by adding parking facilities or, in contrast, reduce the availability of parking spaces in order to discourage drivers from entering the area. Our method can be applied for predicting changes in parking search times and the overall area affected by the proposed changes.

It is very natural that parking studies, including ours, focus on the estimation of system parameters in the situation when parking supply is insufficient and drivers can cruise for lengthy periods. However, no matter how deep the knowledge regarding parking search times, the goal of a planner is to improve the traffic in city centers [Marsden, 2006]. A self-evident and highly unpopular solution here is to reduce the number of cars that enter the city center, and this can be achieved in different ways. Besides rigid direct limitations on vehicular entrance to a designated area, policy makers can also decrease the demand by increasing parking prices [Shoup, 2006; Gragera and Albalate, 2016; Pierce and Shoup, 2013; SFMTA, 2016; Chatman and Manville, 2014; Millard-ball et al., 2014; Cats et al., 2016; Alemi et al., 2018]. Furthermore, parking prices can be flexible, adapting to the fluctuations in the demand for parking throughout the day. Incorporation of cruising drivers' reactions to parking prices would extend our approach, and the first step in this direction is presented in [Fulman and Benenson, 2019].